# On-chip room-temperature CW lasing from a III-V nanowire integrated with a Si photonic crystal platform

**MASATO TAKIGUCHI,**[1,2,*] **TAKURO FUJII,**[1,3] **HISASHI SUMIKURA**[1,2], **AKIHIKO SHINYA**[1,2], **SHINJI MATSUO,**[1,3] **AND MASAYA NOTOMI**[1,2,4]

[1]*NTT Nanophotonics Center, NTT, Inc., 3-1 Morinosato Wakamiya, Atsugi, Kanagawa 243–0198, Japan*
[2]*Basic Research Laboratories, NTT, Inc., 3-1 Morinosato Wakamiya, Atsugi, Kanagawa 243–0198, Japan*
[3] *Device Technology Laboratories, NTT, Inc., 3-1 Morinosato Wakamiya, Atsugi, Kanagawa 243–0198, Japan*
[4]*Department of Physics, Tokyo Institute of Science, 2-12-1 Ookayama, Meguro-ku, Tokyo 152–8550, Japan*

**masato.takiguchi@ntt.com*

**Abstract:** We report the demonstration of continuous-wave (CW) lasing at room temperature from a III-V semiconductor nanowire integrated into a Si photonic crystal (PhC) cavity. Previous hybrid nanowire lasers [M. Takiguchi. et.al., APL Photonics, 2, 046106 (2017)], which typically feature circular nanowire-cross-sections, suffer from a weak optical confinement, preventing CW lasing under ambient conditions. To overcome this limitation, we fabricated nanowires with rectangular cross-sections via dry etching and integrated them into the air trenches of Si PhC cavities formed using atomic force microscope tips. This configuration forms a hybrid photonic crystal cavity with an improved optical confinement. As a result, we achieved room-temperature CW oscillation from a single nanowire, representing a significant step toward on-chip nanophotonic light sources. This unique in-plane integration of the nanolaser in the same plane as the Si slab rather than on top of the substrate will contribute to the development of compact, scalable, and CMOS-compatible photonic circuits.

## 1. Introduction

The convergence of photonics and electronics [1] alongside advancements in on-chip optical computing [2] [3] [4] [5] has attracted significant research attention in recent years. There are several challenges facing the practical implementation of photonic computing, including the high-density integration of photonic nanodevices and the enhancement of individual device performance through advanced nanophotonic structures. To address these challenges, we aim to develop ultra-compact photonic devices utilizing nanophotonic technologies such as photonic crystals (PhCs). These devices are expected to require substantially lower operating energy compared to their conventional counterparts. A key technological milestone in this field is the integration of other materials (especially III-V semiconductor) into Si platforms, which is essential for future optoelectronic integrated circuits [6]. The fundamental components for photonic computing systems are primarily photonic nanodevices such as nanolasers, photodetectors, and optical switches. In particular, the integration of nanolasers into silicon chips is essential for making heterogeneous integration schemes a reality. It will also be necessary to develop continuous-wave (CW) nanolasers that can operate in the telecommunication wavelength range.

Among the various integration techniques proposed thus far, we focus in this study on the integration of III–V nanowires, which feature compact dimensions alongside promising optical and electrical properties [7] [8] [9] [10] [11]. These nanowires can serve as efficient light emitters, enabling the development of hybrid systems that combine III–V materials with Si substrates, thereby paving the way toward energy-efficient photonic components. Nanowire lasers have been demonstrated across a wide range of wavelengths, among which devices operating at telecommunication wavelengths are particularly important for integrated photonics [7] [12] [13] [14]. Research groups have reported the implementation of such lasers utilizing either single nanowires [15] [16] or periodic nanowire arrays as gain media [17] [18] [19]. However, single nanowire lasers face significant limitations, including poor optical confinement due to low cavity quality factors, large mode volumes, and high lasing thresholds [20]. As a result, most operate only under pulsed excitation, and achieving CW lasing at room temperature from a single nanowire remains challenging. Previous studies have indicated that room-temperature CW lasing requires the threshold to be reduced by at least one order of magnitude. This reduction necessitates the introduction of a high-Q cavity to enhance the light-matter interaction. In addition to these optical challenges, there are also practical issues related to lateral integration. Since on-chip optical signals typically propagate through planar waveguides, vertically standing nanowires are often incompatible with efficient in-plane coupling, which hinders their integration into photonic circuits.

Several strategies have been proposed to address these limitations, particularly with respect to cavity Q-factors and optical confinement. One approach involves directly incorporating nanocavities into individual nanowires [21] [22]. While this method can potentially improve the confinement, it typically requires nanowires with larger diameters and involves complex fabrication processes, such as focused ion beam milling or patterning with dielectric hard masks. These methods are often constrained by fabrication precision and scalability. A second strategy employs periodic nanowire array structures [23] [24]. This approach mitigates the influence of air gaps between the nanowire and surrounding medium and is more compatible with future monolithic integration schemes. Our investigations confirm that such nanowire array cavities can achieve high Q-factors through inverse design methodologies. However, despite these benefits, the design flexibility remains somewhat limited due to geometric constraints imposed by the array configuration. A third and highly promising strategy involves the hybrid integration of nanowires with PhC cavities [25] [26] [27]. PhCs are known for their exceptional optical properties, including ultra-high quality (Q) factors and ultra-small mode volumes, making them highly suitable for low-threshold photonic devices. Additionally, Si PhCs are fully compatible with CMOS fabrication processes and allow for lateral device integration. However, Si is an indirect bandgap semiconductor, which makes it inefficient as a light emitter. In contrast, as previously discussed, III-V semiconductor nanowires exhibit direct bandgap emission, along with other advantageous properties such as low parasitic capacitance [28]. The latter is particularly beneficial for future high-speed optical signal modulation owing to the reduced RC time constants. Nevertheless, nanowires alone typically lack high-performance optical cavities due to their geometrical and material constraints. To address this mismatch, we employ precise nanomanipulation techniques—such as AFM [25] [29], microneedles/microgrippers [20] [30] [31], and transfer printing [32] [33] [34] [35]—to physically integrate nanowires with Si PhC cavities. The resulting hybrid platform leverages the complementary strengths of both components: efficient light emission and electrical performance from nanowires, and high-Q optical confinement from PhCs. We believe this hybrid integration approach offers a promising solution for high-performance optoelectronic devices in photonics-electronics convergence.

The main advantages of our hybrid PhC–nanowire system are as follows:

(i) By preparing multiple nanowires with different functionalities and compositions, various photonic devices can be integrated on the same Si platform.

(ii) Functional elements such as heterostructures, multiple quantum wells (MQWs), quantum dots, and p–n junctions can be incorporated inside the nanowires, allowing these ultra-compact functional elements to be utilized within the PhC platform (or Si platform).
(iii) Although nanowires are generally considered unsuitable for in-plane integration, hybridization with PhCs enables waveguide connections and inter-device coupling.
(iv) While individual nanowires typically suffer from insufficient optical confinement, thereby necessitating large diameters to achieve adequate performance, combining them with PhCs allows for stronger optical confinement, and functional devices with subwavelength-sized nanowires can be achieved.

Indeed, we have demonstrated a variety of photonic devices using this hybrid system, including lasers [36], optical switches [37], and photodetectors [28].

Despite these advances, several challenges remain. One notable issue is related to thermal management: due to their high thermal resistance and geometrical characteristics (e.g., cylindrical shape and point contact with the substrate), nanowires are prone to heat accumulation under optical pumping. Moreover, self-catalytically grown nanowires often exhibit non-uniform, tapered structures. These geometrical asymmetries can lead to undesirable optical field distribution, where part of the electromagnetic field becomes trapped in air gaps surrounding the nanowire, thereby significantly reducing the quantum efficiency. As a result, we have only observed lasing in our current hybrid systems under cryogenic conditions [38]. Overcoming these thermal and structural limitations is essential for achieving room-temperature operation and further advancing the performance and practicality of nanowire-based photonic devices.

In the current study, we successfully eliminated undesirable air gaps by precisely integrating etched nanowires into Si PhC L3 cavities (Fig. 1(a) and (b)). We first conducted numerical simulations to evaluate how the inevitable air gaps introduced during fabrication affect the cavity Q and optical mode confinement. Then, on the basis of these insights, we systematically optimized the placement and geometry of the air gaps to identify configurations that preserve high Q factors and strong light confinement. We next experimentally implemented both dry and wet etching techniques to fabricate nanowires with tailored geometries, and subsequently integrated them into pre-formed air trenches within Si PhC cavities using a combination of transfer printing and AFM-based nanomanipulation. This integration strategy enabled precise spatial alignment and strong coupling between the nanowire and the optical cavity. As a result, we achieved significant enhancement in optical confinement and achieved a high-Q cavity that supports CW lasing from a single nanowire at room temperature.

## 2. Design

Our primary objective is to suppress optical leakage caused by local field enhancement at material discontinuities [39] (see Fig. S1(a) in Supplement 1). In dielectric nanostructures, electric fields naturally concentrate near sharp interfaces and air gaps due to the discontinuity in dielectric permittivity. While this phenomenon can enhance light–matter interaction when the emitter is located in air (low-index side), in nanowire-PhC hybrid systems it often leads to significant confinement loss, particularly when electric field maxima coincide with low-index air regions. We therefore propose a cavity design in which semiconductor nanowires with rectangular cross-sections are embedded into the etched trenches of a Si PhC cavity. Our previous hybrid nanowire lasers [38] typically employ nanowires with circular cross-sections, which result in non-uniform contact with the substrate and form inevitable air gaps (see Fig. S1(b) in Supplement 1). These air gaps, located in regions of strong electric field, lead to poor optical confinement and increased radiative losses. In contrast, rectangular nanowires provide planar interfaces that reduce unintended air gaps and enable better physical contact with the surrounding Si structure. This geometry enhances thermal conduction and also mitigates the local field enhancement-induced leakage by suppressing field localization in the air region.

We systematically analyzed the impact of $\Delta x$ and $\Delta y$ gaps introduced during the fabrication process using the finite element method, as illustrated in Fig. 2(a). In our simulations, the lattice constant of the Si PhC was set to 400 nm, the hole radius to 200 nm, the slab thickness to 220 nm, the trench depth to 160 nm, the trench length to 1200 nm, and the trench width to 400 nm. A perfect electric conductor boundary was placed at the center of the L3 cavity to reduce the computational cost. Figure 2(b) shows the electric field distributions of the fundamental mode for three cases: no gap, $\Delta x = 50$ nm ($\Delta y = 0$), and $\Delta y = 25$ nm ($\Delta x = 0$). In the gapless case, a typical L3 cavity mode profile is observed, while in the two cases with gaps, the electric field is pulled toward the gap region (see Fig. S2 in Supplement 1 for the simulation results with higher modes).

Figures 2(c) and (d) show the systematic dependence of the cavity Q-factor and radiation rate $\Gamma$ on the gap dimensions. These results show that $\Delta y$ has a strong influence on both Q and $\Gamma$. We attribute this to the dominant Ey component in the cavity mode and the fact that the vertical slot intersects the field, meaning that even a small $\Delta y$ gap can significantly degrade the Q-factor and increase the radiation loss (see Fig. S1(c) and (d) in Supplement 1). On the other hand, the $\Delta x$ gap, which coincides with an electric field node, has a relatively minor effect. This is because the slot is aligned parallel to the cavity mode (polarization), and its position corresponds to the field node.

In practical device fabrication, it is challenging (though not entirely impossible) to insert a nanowire into the slot without introducing either a $\Delta x$ gap or $\Delta y$ gap. However, these simulation results suggest that allowing a finite $\Delta x$ is acceptable, as its impact is limited. On the basis of these findings, our design guidelines are to (1) minimize the $\Delta y$ gap, ideally achieving near-gapless contact to prevent field penetration into the air, and (2) position any unavoidable $\Delta x$ gaps at the electric field nodes of the cavity mode, where the field amplitude is minimal.

## 3. Fabrication

The fabrication process (Fig. S3 in Supplement 1) of the nanowire– PhC hybrid device begins with the definition of nanowire patterns on an InP wafer containing InGaAsP multiple quantum wells (MQWs). We utilize electron-beam lithography to define the nanowire geometry, followed by dry etching to achieve vertical sidewalls. Subsequently, the InP substrate beneath the nanowires is selectively removed using a wet chemical etching process, resulting in suspended nanowires (Fig. 3(a)). A transfer process is then implemented to integrate these nanowires into the Si PhC substrate. First, nanowires are picked up from the InP substrate using a polydimethylsiloxane (PDMS) stamp and transferred onto the Si PhC substrate via contact stamping. In cases where multiple nanowires are transferred simultaneously, individual nanowires are selectively manipulated using a micromanipulator to position one with the desired length near the PhC cavity. When a single nanowire is successfully transferred during stamping, this intermediate step can be omitted. For final placement, AFM-based nanomanipulation is utilized to accurately align and insert the nanowires into the predefined PhC trenches. The nanowires are gently pushed downward using the AFM probe to ensure intimate contact with the substrate.

Figure 4(a) shows an SEM image of the PhC prior to the introduction of the nanowire, and Figs. 1(b), 4(b), and 4(c) display SEM images taken after the fabrication process described in Fig. S3 in Supplement 1. The width of the nanowires is 400 nm in the former case and 200 nm in the latter. As we can see, there is no visible vertical gap ($\Delta y$ gap) between the nanowire and the substrate. Furthermore, the nanowires lie flat with minimal protrusion, indicating good contact with the substrate. Although minor fabrication imperfections (such as edge roughness or surface burrs) are observed, these can be mitigated by optimizing the etching conditions. Although not shown here, we also successfully integrated rectangular nanowires with widths as small as 200 nm into the trenches. However, as discussed later, these did not exhibit lasing.

Due to the small cross-sectional area of the nanowires, surface recombination may significantly influence the carrier dynamics. To evaluate this effect, we performed time-resolved photoluminescence (TRPL) measurements to determine the carrier lifetime in nanowires with varying widths (see Fig. S4 in Supplement 1). We compared nanowires with widths ranging from 100 nm to 600 nm and the as-grown MQW wafer. As shown Fig. 3(b) and (c), the results show a clear trend: narrower nanowires exhibit shorter carrier lifetimes, indicating an increased surface recombination rate. While this fast non-radiative process could be detrimental to lasing by raising the threshold, it can potentially be addressed through crystal regrowth techniques, which enable the formation of nanowires with passivating InP cap layers [40] [41]. For example, previously fabricated thin self-catalytic InP nanowires with MQWs and InP cap layers have demonstrated significantly longer lifetimes [36]. Moreover, even when nanowires are fabricated by dry etching, it is possible to produce nanowires with InP capping layers by employing an embedding regrowth technique commonly used in InP-based photonic crystal fabrication [42] [43]. In this approach, the nanowire structure containing MQWs is first embedded in InP through regrowth, and subsequently, the nanowire geometry is defined again by a second dry etching process. We should also point out that short-lifetime structures can be advantageous for other applications, such as high-speed all-optical switching, where rapid carrier recombination is desirable for faster response times [44].

## 4. Experiment

We performed photoluminescence (PL) measurements on the fabricated samples using a nanosecond pulsed laser as the excitation source. To avoid degradation of the Q-factor, we employed isolated cavities without integrated waveguides, as shown in Fig. 5(a). The use of pulsed excitation minimizes sample heating and provides optimal conditions for evaluating the lasing threshold and emission characteristics. We measured and compered four types of samples. Three of them consisted of nanowires with widths of 250 nm, 300 nm, and 400 nm embedded into PhC trenches of matching width, thereby eliminating any $\Delta y$ air gaps. The fourth sample consisted of a 300-nm-wide nanowire embedded into a 400-nm-wide trench, intentionally introducing a $\Delta y$ air gap.

Figure 5(b) shows the PL spectra of a 400-nm-wide nanowire under strong and weak excitation conditions. As we can see, a distinct and sharp emission peak is observed at $\lambda = 1565$ nm under strong excitation. To confirm the reproducibility of lasing across different samples, the Supplementary Information presents the experimental results obtained for nanowires with different widths and lattice constants (see Fig. S5 in Supplement 1). As shown in the light-in versus light-out (L-L) plots in Fig 5(c), both gapless samples exhibited clear lasing behavior, whereas the gapped sample failed to achieve lasing under the same excitation conditions. This result highlights the importance of minimizing the air gap in the $\Delta y$ direction to preserve high optical confinement in the cavity. The typical lasing threshold was approximately 0.15 mJ cm$^{-2}$/pulse.

We also compared the performance of the new hybrid laser with that of a previously reported single nanowire laser fabricated by our group [20] and measured under identical conditions. The lasing threshold of the simple nanowire laser at this time was 2.1 mJ cm$^{-2}$/pulse. Despite the increased non-radiative surface recombination expected in the newly etched nanowires, the hybrid laser exhibited a lasing threshold that was reduced by more than one order of magnitude. This significant improvement demonstrates the superior optical confinement provided by the gapless PhC cavity.

To further quantify the influence of nanowire geometry on lasing performance, we conducted a rate equation analysis assuming a two-level system with linear gain characteristics (see Fig. S6 in Supplement 1) [45]. To support this analysis, we performed additional TRPL measurements (see Fig. S7 in Supplement 1). The results indicate that the emission within the nanocavity is dominated by fast non-radiative recombination processes, and no observable Purcell enhancement was detected. As shown Fig. 5(d), experimental L–L curves for nanowires with varying widths were analyzed, and the threshold values were normalized to the 400-nm-wide nanowire condition. In addition to the data shown Fig. 5(d), we measured a larger number of nanowire lasers and incorporated these results into the updated figure. For each nanowire width, at least three devices were characterized, and error bars were added to reflect the measurement variability. These additional results are also included in the Supplementary Information (Fig. S5 in Supplement 1). The results indicate a sharp increase in threshold as the nanowire width decreases, consistent with enhanced non-radiative recombination due to increased surface-to-volume ratio. Notably, nanowires with a width of 200 nm failed to exhibit lasing, which can possibly be attributed to gain saturation and thermal effects occurring prior to achieving threshold.

We also performed CW excitation experiments under a room-temperature condition. As shown Fig. 6(a), bright near-infrared emission was observed via IR camera imaging when the pump power is strong. The Spectra, L-L plot and the linewidth dependence on pump intensity are shown in Fig. 6(b) and (c). The L–L curve revealed a characteristic linewidth narrowing with increasing emission intensity, confirming lasing action. Based on the linear region of the L–L plot after the onset of lasing, the threshold was estimated to be 40 µW (Power density: 10 kW/cm$^2$). This lasing threshold is comparable to that of other photonic crystal lasers operating under optical pumping in continuous-wave mode at room temperature [46] [47] [48], and it can potentially be further improved through future design optimization. Finally, we evaluated the photon statistics of the laser emission using a Hanbury Brown and Twiss interferometer setup (see Fig. S4(b) in Supplement 1). In general, lasing from nanolasers is often ambiguous, and analyzing the photon statistics provides strong evidence of lasing. In the present case, although the threshold is clearly observed because of the significant nonradiative recombination, we performed the measurements for confirmation. Near the threshold, where spontaneous emission dominates, the second-order correlation function $g^{(2)}(0)$ exhibited clear bunching behavior (Fig. 6(d)). Additionally, sinusoidal oscillations on both sides of the central peak, attributed to relaxation oscillations, were observed [38] [49] [50]. At higher excitation powers, $g^{(2)}(0)$ approached unity, indicating a transition to coherent light emission. From this series of experiments, we successfully demonstrated that our subwavelength nanowire laser achieved CW operation at room temperature.

## 5. Conclusion and outlook

In this study, we demonstrated the successful implementation of subwavelength-scale III–V nanowires in pre-defined trenches within Si PhC cavities with nanometer precision. This hybrid integration strategy enabled strong optical confinement and resulted in clear lasing oscillation under both pulsed (nanosecond and picosecond) and CW excitation conditions. Notably, CW lasing was achieved at room temperature using subwavelength nanowires, which is an important step in the development of energy-efficient, chip-scale light sources. This planar embedding of the nanowire laser within the Si photonic layer, rather than mounting it on top, represents a vital step toward the seamless integration of active light sources into standard silicon photonic platforms. Furthermore, our results revealed that the lasing threshold is strongly influenced by non-radiative recombination processes, particularly surface recombination, which becomes more pronounced as the nanowire dimensions are reduced. This behavior was validated through both experimental measurements and theoretical modeling, establishing an important design consideration for future device optimization. Our work also provides design guidelines [51] [52] that are directly applicable to future efforts toward selective epitaxial growth of nanowires on Si substrates. By addressing challenges related to optical confinement and non-radiative losses, our approach contributes to the scalable integration of nanowire lasers in silicon photonic platforms. Looking ahead, further improvements can be envisioned through the introduction of nanowire regrowth techniques aimed at suppressing surface recombination, thereby reducing the lasing threshold and improving device stability. Additionally, future developments will explore extension to visible wavelength operation [53], the development of current-injection nanowire lasers [23] [54] [55], multi-wavelength integration for photonic circuit complexity [56], nonlinear devices beyond lasers [57] (such as all-optical switches [37] and memory [58]), and implementation of coupled resonator architectures [59] to explore novel lasing regimes and sensing applications. These advancements will contribute to the broader goal of integrating active III–V nanophotonic elements with silicon-based platforms, accelerating the fusion of photonics and electronics at the chip scale.

**Funding**

This work has been supported by the JSPS KAKENHI Grant Numbers 15H05735, 21H01834, and 23K26581.

**Acknowledgments**

The authors express their sincere gratitude to Shinichi Fujiura for his invaluable assistance with the device measurements and nanomanipulation. They also acknowledge Yoshio Ohki, Toshifumi Watanabe, and Osamu Moriwaki of NTT Advanced Technology Corporation, as well as Junichi Asaoka of NTT Devices Cross Technologies Corporation, for their expert technical support in the SEM observation, electron-beam lithography, and fabrication processes of both Si and InP devices.

**Disclosures**

The authors declare no conflicts of interest.

**Data availability**

Data underlying the results presented in this paper are not publicly available at this time but may be obtained from the authors upon reasonable request.

**Supplemental document**

See Supplement 1 for supporting content.

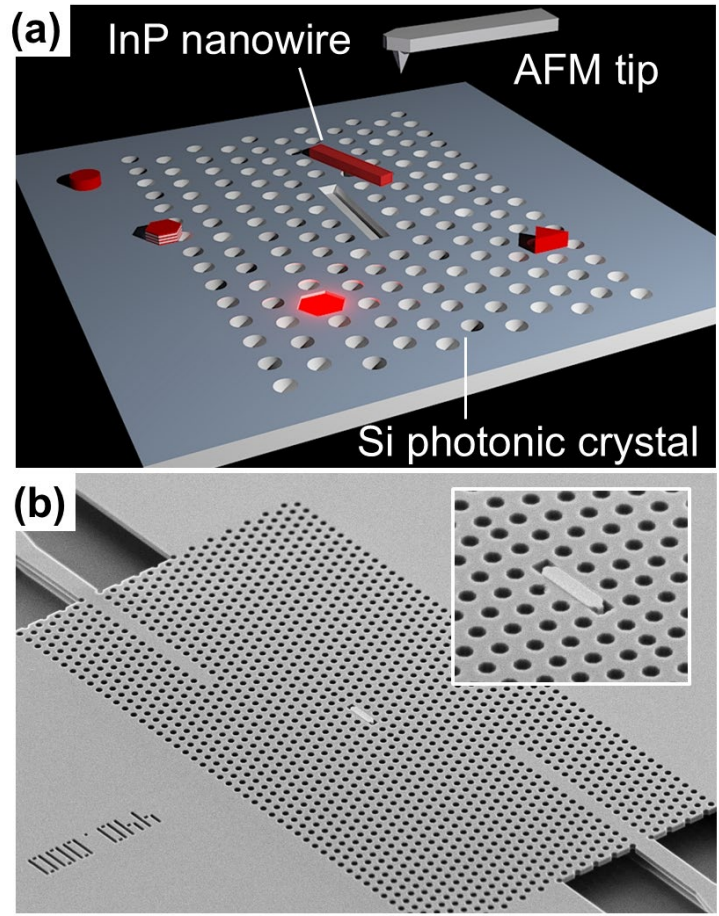

Fig. 1. (a) Schematic illustration of Si photonic crystal and III-V nanowire manipulated using an atomic force microscopy tip. (b) Angled scanning electron microscope image of III-V nanowire (width: 250 nm) integrated into the L3 cavity of a Si photonic crystal with a lattice constant of 440 nm and hole diameter of 200 nm. Inset shows a magnified view of the nanowire.

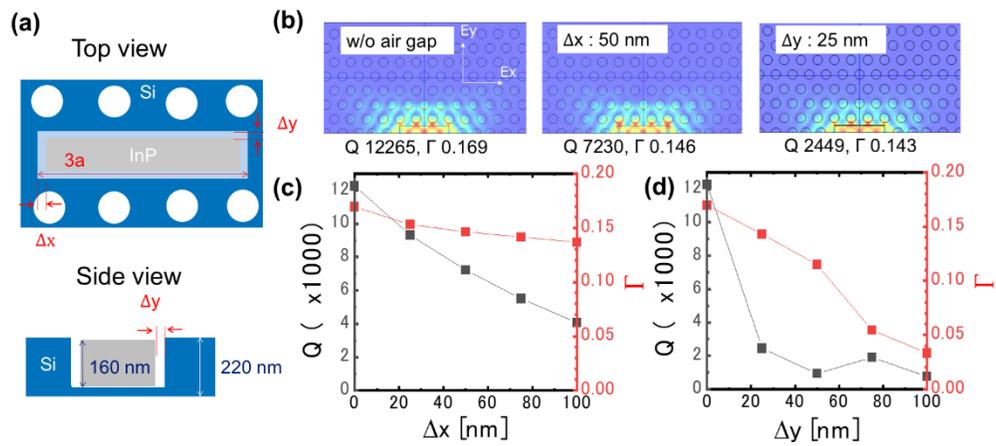

Fig. 2. (a) Schematic illustrations of nanowire and photonic crystal (top and side views). (b) Electric field distribution of photonic crystal cavity (no gap, $\Delta x = 50$ nm, and $\Delta y = 25$ nm). (c) Relationship between Q and $\Gamma$ as a function of $\Delta x$. (d) Relationship between Q and $\Gamma$ as a function of $\Delta y$.

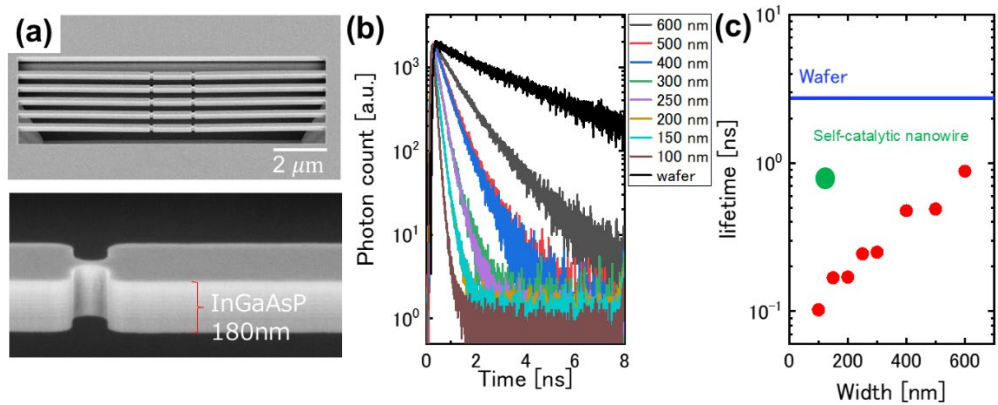

Fig. 3. (a) Angled scanning electron microscope images of suspended InP nanowires. (b) Emission lifetime of nanowires and wafer. (c) Nanowire width vs. emission lifetime for different nanowires.

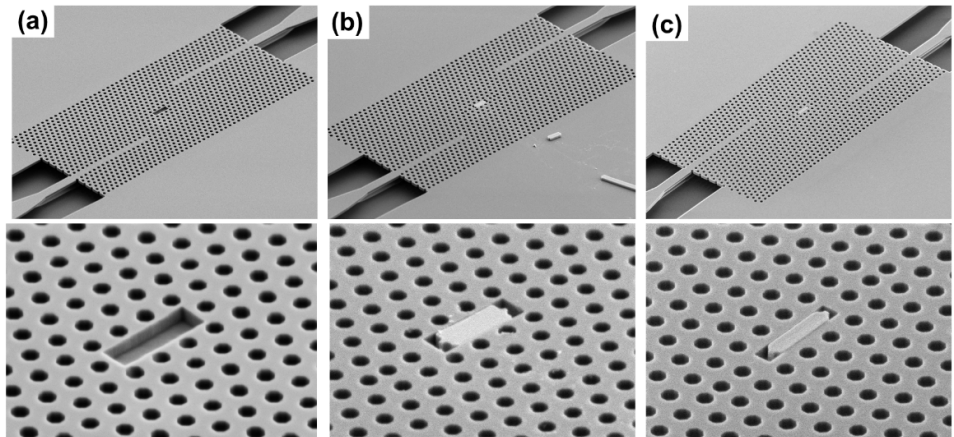

Fig. 4. (a) SEM image without nanowire. (b) SEM image with nanowire (width: 400 nm). (c) SEM image with nanowire (width: 250 nm).

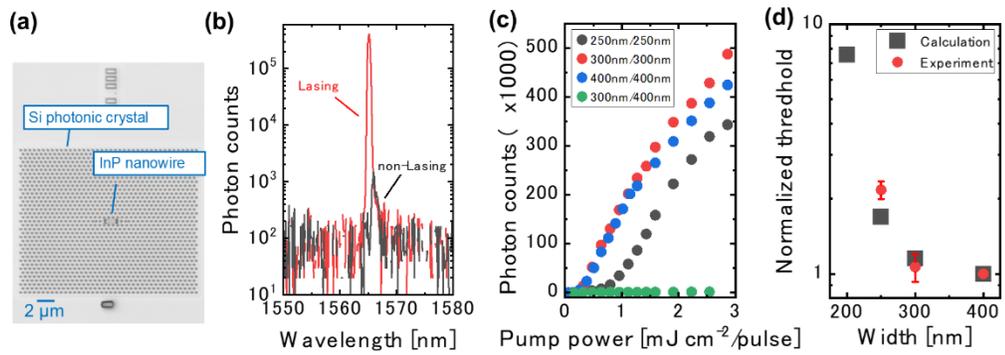

Fig. 5. (a) Confocal laser scanning microscope image of L3 cavity with nanowire (lattice constant of 430 nm and nanowire/trench width of 400 nm/400 nm). (b) PL spectra before and after laser oscillation for the sample. (c) L-L curves at different samples. Each sample has a nanowire width/slot width of 250 nm/250 nm (black dots), 300 nm/300 nm (red dots), 400 nm/400 nm (blue dots), and 300 nm/400 nm (green dots). (d) Nanowire width vs. lasing thresholds for different nanowires. Error bars show the standard deviation.

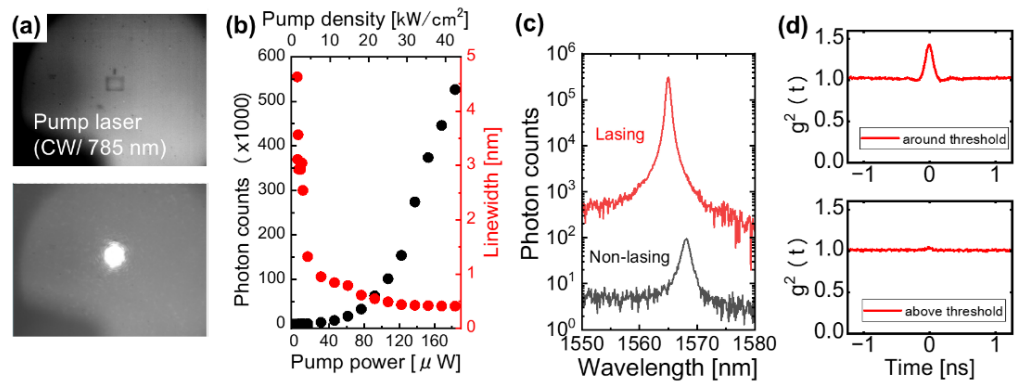

Fig. 6. (a) IR image from the top under strong pumping conditions (bottom) and without pumping (top). (b) L-L curve and cavity linewidth. The lower x-axis corresponds to the effective power absorbed by the nanowire, while the upper x-axis corresponds to the power density. (c) Emission spectra. (d) Photon correlation measurement for different pump powers.

# Supplemental document: On-chip room-temperature CW lasing from a III-V nanowire integrated with a Si photonic crystal platform

**Masato Takiguchi,**[1,2,*] **Takuro Fujii,**[1,3] **Hisashi Sumikura**[1,2], **Akihiko Shinya**[1,2], **Shinji Matsuo,**[1,3] **and Masaya Notomi**[1,2,4]

[1]*NTT Nanophotonics Center, NTT, Inc., 3-1 Morinosato Wakamiya, Atsugi, Kanagawa 243–0198, Japan*
[2]*Basic Research Laboratories, NTT, Inc., 3-1 Morinosato Wakamiya, Atsugi, Kanagawa 243–0198, Japan*
[3] *Device Technology Laboratories, NTT, Inc., 3-1 Morinosato Wakamiya, Atsugi, Kanagawa 243–0198, Japan*
[4]*Department of Physics, Tokyo Institute of Science, 2-12-1 Ookayama, Meguro-ku, Tokyo 152–8550, Japan*

[*]*masato.takiguchi@ntt.com*

## S1: Electric field distribution in a slot cavity

Figure S1(a) shows an L3 cavity structure with an embedded slot. In this configuration, the electric field is strongly concentrated in the air region. The fundamental mode of the L3 cavity is predominantly composed of the Ey component, which is oriented perpendicular to the slot. Figure S1(b) illustrates the electric field distribution when a cylindrical InP nanowire is integrated into the trench within a silicon photonic crystal waveguide. We can see here that the electric field is concentrated at the air gaps and at the interfaces between the nanowire and the silicon. This field localization reduces optical confinement within the nanowire, thereby degrading the quantum efficiency of the device. Figure S1(c) and (d) present the electric field distributions for L3 cavities with slots placed at the cavity center in different orientations. In Fig. S1(c), the trench is oriented perpendicular to the dominant Ey component. This configuration significantly perturbs the cavity's field distribution. In contrast, the configuration in Fig. S1(d), where the trench is aligned parallel to the Ey component, has a minimal impact on the field distribution.

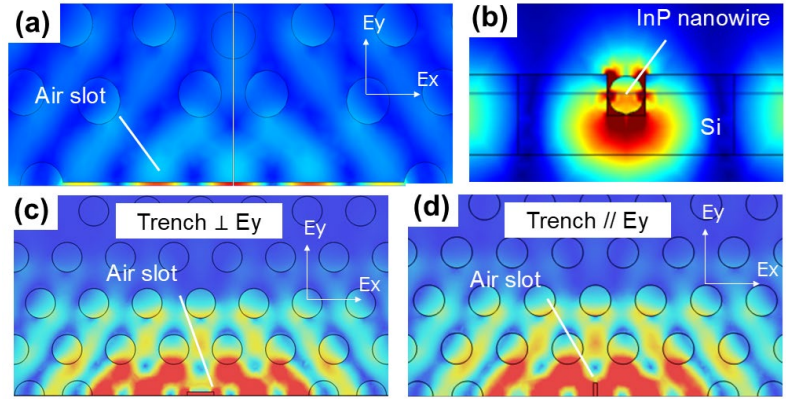

Fig. S1. (a) Cavity mode profile of L3 cavity with an air slot. (b) Cross-section of electric field of the resonator mode when a cylindrical nanowire is inserted into a photonic crystal waveguide. (c) Electric field distribution of the resonator mode when a trench is placed perpendicular to Ey in the L3 cavity. (d) Electric field distribution of the cavity mode when a trench is placed parallel to Ey in the L3 cavity.

## S2: ELECTRIC FIELD DISTRIBUTION IN L3 CAVITY WITH NANOWIRE

The electric field distribution of a silicon L3 cavity with an introduced trench and an integrated rectangular-cross-section InP nanowire is shown in Fig. S2, where (a) presents the fundamental mode, along with the quality factor (Q) and confinement factor ($\Gamma$) for configurations with and without an air gap, and (b) shows the higher-order mode. The higher-order mode cannot be utilized in our experiment due to its intrinsically low Q.

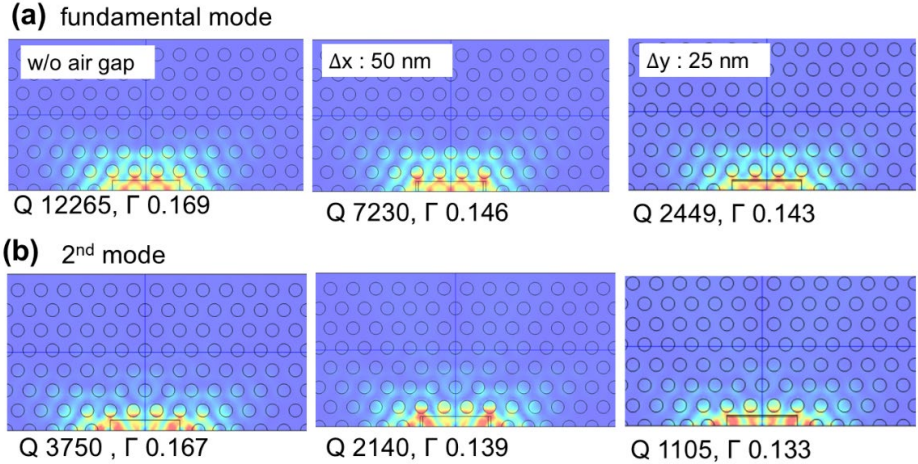

Fig. S2. (a) Cavity mode profile of L3 cavity with an air slot. (b) Cross-section of electric field of the resonator mode when a cylindrical nanowire is inserted into a photonic crystal waveguide.

## S3: FABRICATION PROCESS

The fabrication process is illustrated in Fig. S3. InP nanowires are fabricated from an InP wafer using electron-beam lithography followed by dry etching. This method allows for the creation of nanowires as well as arbitrary geometries beyond simple wire shapes. Subsequently, the structures are suspended via wet etching to form air bridges. To control the final length of the nanowires, predefined notches are introduced during the design phase. In the CAD layout, these notched regions are designed to be only several tens of nanometers wide, making them mechanically fragile and easy to break. The air-bridged nanowires are transferred from the InP wafer to a silicon substrate using a transfer printing technique. For the transfer, we utilize PDMS stamps with surface relief structures on the order of tens of micrometers. When multiple nanowires are transferred simultaneously, specific ones are selected using microgrippers or microneedles and positioned near the photonic crystal cavity. In cases where a single nanowire is selected with PDMS, it is directly transferred in proximity to the photonic crystal. After placement, the transferred nanowires are imaged using atomic force microscopy (AFM) in tapping mode. To position the nanowire precisely, the AFM is then operated in contact mode to laterally push the nanowire toward the target region. This procedure is repeated to guide the nanowire into the trench of the photonic crystal. Once the nanowire is successfully introduced into the cavity region, we use tapping-mode AFM to apply pressure from above, improving the contact and mechanical stability.

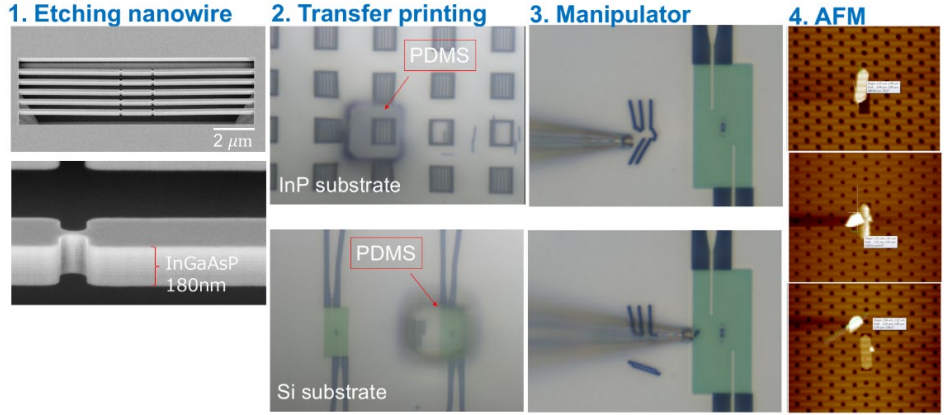

Fig. S3. Fabrication process.

## S4: Measurement setup

Figure S4(a) shows the schematic setup used for the photoluminescence (PL) and time-resolved measurements. The excitation source is varied depending on the measurement. For the pulsed excitation experiments, a laser with a wavelength of 1064 nm, a repetition rate of 250 kHz, and a pulse width of 10 ns is used. For the time-resolved photoluminescence (TRPL) measurements, a femtosecond laser operating at 880 nm with a repetition rate of 80 MHz is used. In the continuous-wave (CW) excitation measurements, a laser with a wavelength of 780 nm is utilized. Emission spectra are acquired using a spectrometer equipped with a liquid-nitrogen-cooled InGaAs detector array. The sample is observed through a 50× objective lens, which also focuses the excitation light onto the sample. Emitted photoluminescence is collected and separated from the excitation light using a long-pass filter. The excitation intensity is controlled using an attenuator, allowing for the investigation of pump power dependence. For the fluorescence lifetime measurements, a superconducting single-photon detector (SSPD) is utilized. Figure S4(b) depicts the optical setup used for the photon correlation measurements.

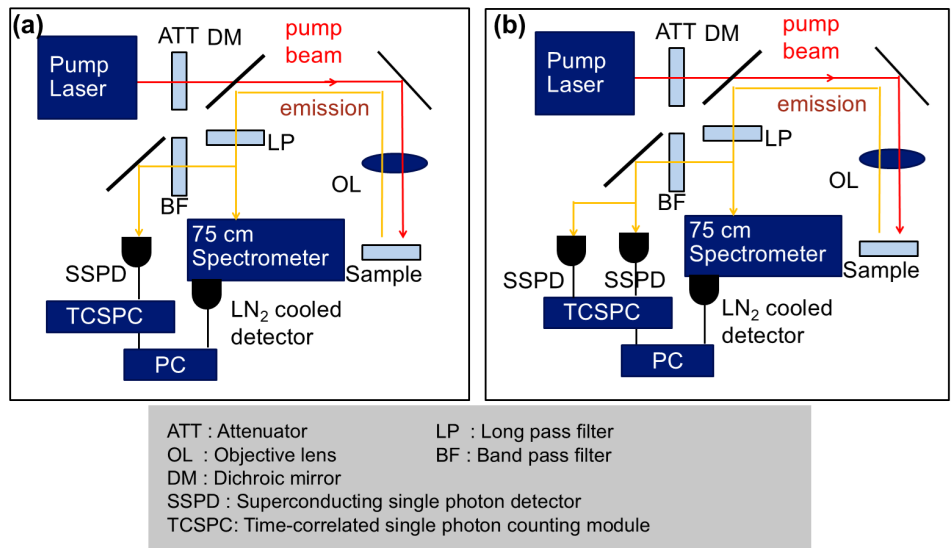

Fig. S4. (a) PL and TRPL measurement setup. (b) Photon correlation measurement setup.

## S5: PL and PL Measurement

We investigated the laser characteristics of several samples. Figure S5 shows the results for samples with nanowire/trench widths of 250 nm/250 nm, 300 nm/300 nm, and 400 nm/400 nm. In each case, three L-L (light-in vs. light-out) curves are plotted. The lattice constant of the photonic crystal ranges from 400 to 430 nm. As we can see in the figure, the sample with a 250-nm width exhibits a slightly higher threshold.

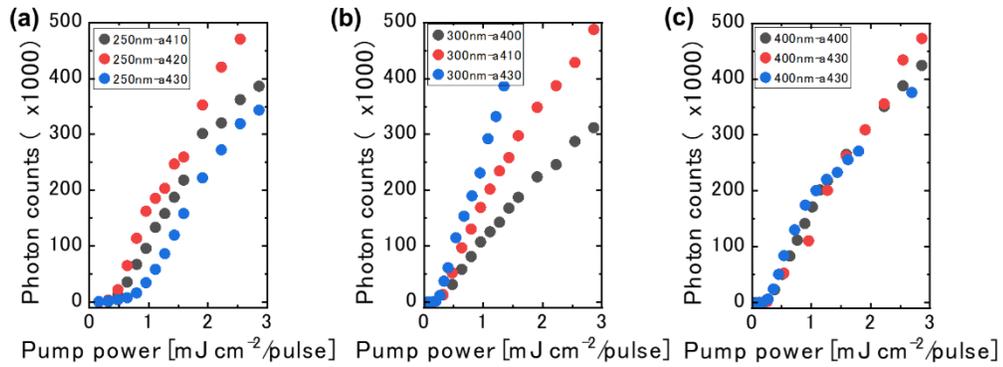

Fig. S5. L-L curves at samples with nanowire/trench widths of (a) 250 nm/250 nm, (b) 300 nm/300 nm, and (c) 400 nm/400 nm.

## S6: TIME-RESOLVED PHOTOLUMINESCENCE MEASUREMENT

We estimated the lasing threshold of the nanowire laser using the rate equations [1] for a typical two-level system as follow. $N$ is the free carrier density in the active medium and $P$ is the photon population in the cavity. In this model, a linear gain is assumed.

$$\frac{dN}{dt} = \frac{I}{qV} - \left(\frac{1}{\tau_{sp}} + \frac{1}{\tau_{nr}}\right)N - \frac{gP}{V}$$

$$\frac{dP}{dt} = -(\gamma - g) + \frac{\beta NV}{\tau_{sp}}$$

$$g = g'(N - N_0)$$

Here, I, $q$, $V$, $\gamma$, $\tau_{sp}$, $\tau_{nr}$, and $\beta$ refer to the injection current, electron charge, volume of the active material (nanowire volume), cavity photon lifetime, spontaneous emission lifetime, nonradiative recombination lifetime, and spontaneous emission coupling factor, respectively. $g$ is the active material gain, described as

$$g' = \frac{\beta V}{\tau_{sp}}\Gamma,$$

where $\Gamma$ is the optical confinement factor. When the width of the nanowire is varied, both the gain volume and the nonradiative recombination rate (see Fig. 3(c) in the main paper) change simultaneously. Figure S6 shows the L-L plots for nanowires lasers with different widths. As we can see, the narrower nanowires exhibit a significantly increased lasing threshold.

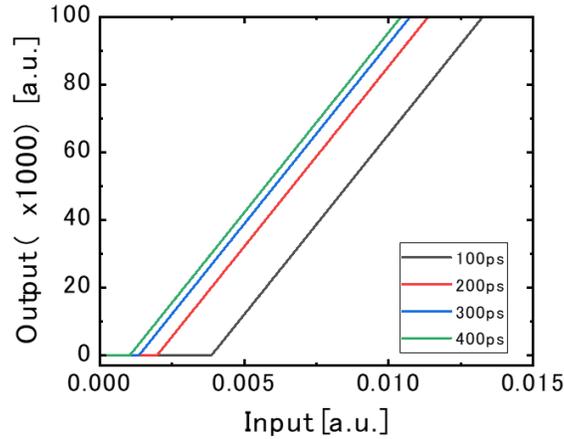

Fig. S6. L-L plots for nanowires lasers with different widths.

[1] G. Björk and Y. Yamamoto, "Analysis of semiconductor microcavity lasers using rate equations," IEEE J. Quantum Electron. 27(11), 2386–2396 (1991).

## S7: TIME-RESOLVED PHOTOLUMINESCENCE MEASUREMENT

In general, systems with a high $Q/V$ ratio, such as nanocavity structures, have the potential to exhibit the Purcell effect. To investigate this possibility, we performed TRPL measurements using a femtosecond laser as the excitation source. Figure S7(a) shows the emission spectra obtained both above and below the lasing threshold. Below the threshold, the measured fluorescence lifetime was approximately 310 ps, which is comparable to that of a nanowire without an optical cavity (Fig. S7(b)). This indicates that no significant acceleration of spontaneous emission is observed under these conditions. The nonradiative recombination processes therefore appears to dominate over the radiative recombination rate, even if the latter were to be enhanced by the Purcell effect. Furthermore, as the excitation intensity was increased, lasing was observed, accompanied by a significant acceleration of the emission due to stimulated emission (Fig. S7(b)). The corresponding decay time reached the temporal resolution limit of the measurement system.

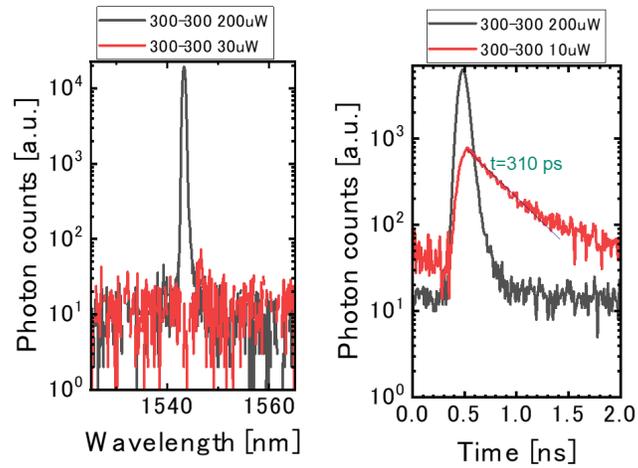

Fig. S7. (a) Cavity mode profile of L3 cavity with an air slot. (b) Cross-section of electric field of the resonator mode when a cylindrical nanowire is inserted into a photonic crystal waveguide.